\documentclass[onecolumn,showpacs,amsmath,amssymb]{revtex4}
\usepackage{enumerate}
\usepackage{dcolumn}
\usepackage{graphicx}
\usepackage{bm}
\begin{document}
\title{Solving the Wheeler-DeWitt of Small Universe}

\author{Shintaro Sawayama}
 \email{sawayama0410@gmail.com}
\affiliation{Sawayama Cram School of Physics\\ Atsuhara 328, Fuji-city, Shizuoka prefecture 419-0201, Japan}
\begin{abstract}
We can solve the Wheeler-DeWitt equation of the small universe enough to metric becomes diagonal 
and take a Gaussian normal coordinate.
Our previous works are concerning to this paper.
In this paper, we only write how to solve the Wheeler-DeWitt equation of such universe.
Our motivation is simple, that is to solve the Wheeler-DeWitt equation.
Even if the Wheeler-DeWitt equation is solved, quantum gravity does not complete yet.
However, this work may be one of the first step to quantum gravity.
\end{abstract}

\maketitle
\section{Introduction}\label{sec1}
In the quantum gravity, there are many approaches, for example loop quantum gravity\cite{As}\cite{Rov}\cite{Thi} or mini-superspace\cite{Hart} 
approachs or string approaches\cite{AL}.
However, quantum gravity has not completed yet.
In the canonical quantum gravity, the difficulties comes from the Wheeler-DeWitt equation.
At first we should solve the Wheeler-DeWitt\cite{De} equation the quantum gravity does not start.

In our previous work is concerning the Wheeler-DeWitt equation.
Our motivation is simple to solve the Wheeler-DeWitt equation.
However, Wheeler-DeWitt is difficult to solve.
Because it is second order elliptic functional partial differential equation with non-linear term.
The elliptic differential equation is difficult, but if we choose diagonal metric,
this problem is solved.
The partial differential equation is difficult, but if we use additional constraint equation,
this problem is solved.
The functional differential equation is difficult, but we construct how to solve the 
functional differential equation.
The non-linear term is difficult, but we construct the method to remove this difficulty.

We quantize diagonal metric universe by using Wheeler-DeWitt without approximation.
We use the fact metric become diagonal by coordinate transformation and Gaussian normal coordinate.
Such universe is small universe.
Or we treat the Wheeler-DeWitt equation locally.
Because gravity is quantize in very small universe.
To treat such small universe is theoretical.

In section \ref{sec2}, we simplify the Wheeler-DeWitt equation. 
In section \ref{sec3} we quantize toy model.
And in this section we show the step to solve the Wheeler-DeWitt.
In section \ref{sec4} we quantize full quantum gravity, using the obtained result in section \ref{sec3}.
In section \ref{sec5} we summarize and conclude obtained result.
\section{Simplification of the Wheeler-DeWitt}\label{sec2}
For simplicity we treat the only diagonal metric universes as,
\begin{eqnarray}
\begin{pmatrix}
g_{00} & 0 & 0 & 0 \\
0 & g_{11} & 0 & 0 \\
0 & 0 & g_{22} & 0 \\
0 & 0 & 0 & g_{33}.
\end{pmatrix}
\end{eqnarray}
And in all the spacetime metrics become diagonal locally.
Or we treat small universe enough to metric become diagonal. 
We start from decomposition of the Einstein Hilbert action of the above diagonal metric universe.
The action of the above universe is written by
\begin{eqnarray}
S=\int RdM=\int R[g_{\mu\mu}]dSdt.
\end{eqnarray}
Here $S$ is the hyper-surface with constant time.
Because, our method is different from the usual Wheeler-DeWitt equation formalism,
our obtained Hamiltonian constraint is a different type of the Wheeler-DeWitt equation.
If we decompose this action as 3+1, then we can obtain
\begin{eqnarray}
{\cal L}=\dot{q}_{ii}P^{ii}+NH-2\sqrt{q}D^iN_{,i}.
\end{eqnarray}
Here $N$ is the lapse functional and $N_{,i}$ is the sift vectors, and $H$ is the Hamiltonian constraint such that
\begin{eqnarray}
H=\frac{1}{2}q_{ii}q_{jj}P^{ii}P^{jj}+{\cal R}.
\end{eqnarray}
Here ${\cal R}$ is the three dimensional Ricci scalar and $P^{ii}$ is the momentum whose commutation relation with $q_{ii}$ is not $i$, it is $i\sqrt{q}$.
In this formulation there are not appear $q_{ij}$ and $P^{ij}$ and sift vectors and momentum constraint.
So we can ignore the constraint as $[P^{ij},H]$ or $[[P^{ij},H],H]$, because we start with metric diagonal setting.
In this simple case we can ignore the diffeomorphism constraints.
If we write the Hamiltonian constraint in the operator representation, we obtain
\begin{eqnarray}
H=\sum_{ij}\frac{1}{2}\frac{\delta^2}{\delta \phi_i\delta \phi_j}+
{\cal R}[q_{11},q_{22},q_{33}]=0 \nonumber \\
=\sum_{ij}\frac{1}{2}\frac{\delta^2}{\delta \phi_i\delta \phi_j}+
\sum_{i\not= j}(\hat{\phi}_{i,jj}+\hat{\phi}_{j,i}\hat{\phi}_{i,i})e^{\hat{\phi}_i}=0.
\end{eqnarray}
Here $\phi_i=\ln q_{ii}$.
If we consider the $\phi$ were only depend $t,x_i$, the Hamiltonian constraint becomes,
\begin{eqnarray}
H= \sum_{ij}\frac{1}{2}\frac{\delta^2}{\delta \phi_i\delta \phi_j}-\Lambda =0.\label{eq1}
\end{eqnarray}
We use this Hamiltonian constraint in section \ref{sec3}
Because we only treat metric diagonal universe, the Hamiltonian constraint has different form from 
the orthodox Hamiltonian constraint.
This setting is similar to mini-superspace model.

If universe is small enough to apply Gaussian normal coordinate,
the Hamiltonian constraint become 
\begin{eqnarray}
H=\sum_{ij}\frac{1}{2}\frac{\delta^2}{\delta \phi_i\delta \phi_j}+
\sum_{i\not= j}\hat{\phi}_{i,jj}e^{\phi_i}=0\label{eq2}
\end{eqnarray}
We treat this Hamiltonian constraint in section \ref{sec4}

The static restriction deviated from up-to-down method\cite{Sa1} is
\begin{eqnarray}
\sum_{i\not=j}\frac{\delta}{\delta\phi_i\delta\phi_j} \label{eq3}.
\end{eqnarray}

\section{Simple example of Wheeler-DeWitt}\label{sec3}
Before solving Eq.(\ref{eq2}), we solve Eq.(\ref{eq1}) for simplicity.
We consider the following spacetime
\begin{eqnarray}
\begin{pmatrix}
-N^2 & 0 & 0 & 0 \\
0 & g_1(t,x) & 0 & 0\\
0 & 0 & g_2(t,y) & 0 \\
0 & 0 & 0 & g_3(t,z) 
\end{pmatrix}.\label{mat1}
\end{eqnarray}
To quantize this spacetime we take steps.
The step 1 is using the static restriction we obtain special solution.
The step 2 is to remove the static restriction we obtain general solution.
The we carry out step 1.
The Eq.(\ref{eq1}) and Eq.(\ref{eq3}) is consistent and simultaneously quantized.
And the solution is 
\begin{eqnarray}
f[\phi_1,\phi_2,\phi_3 ] =\prod_i\exp (a_i\Lambda^{1/2}\int \delta \phi_i)
\end{eqnarray}
where 
\begin{eqnarray}
a_1a_2+a_1a_3+a_2a_3=1 \\
a_1^2+a_2^2+a_3^2=1
\end{eqnarray}

Then we carry out step 2.
Using the above solution as a special solution, we assume the state is a form
\begin{eqnarray}
|\Psi\rangle =f[\phi_i ]g[\phi_i] .
\end{eqnarray}
And we remove the static restriction, then $g[\phi_i]$ should satisfy
\begin{eqnarray}
\nabla (\nabla +a)g[\phi_i]=0
\end{eqnarray}
Here, $\nabla$ is defined by
\begin{eqnarray}
\nabla=\sum_i\frac{\delta}{\delta \phi_i}.
\end{eqnarray}
And $a$ is defined by $a=\Lambda^{1/2}\sum_i a_i$
This equation can be solved easily and solution is 
\begin{eqnarray}
e^{-a\int \delta \phi_1}(-2\int \delta \phi_1+\int \delta \phi_2+\int \delta \phi_3 )
+e^{-a\int \delta \phi_2}(-2\int \delta \phi_2+\int \delta \phi_1+\int \delta \phi_3 ) \nonumber \\
+e^{-a\int \delta \phi_3}(-2\int \delta \phi_3+\int \delta \phi_2+\int \delta \phi_1 )
\end{eqnarray}
So the quantum state of (\ref{mat1}) is
\begin{eqnarray}
|\Psi (\phi_i)\rangle =\prod_i \exp (a_i\Lambda^{1/2}\int \delta \phi_i) 
 \bigg[ e^{-a\int \delta\phi_1}(-2\int \delta \phi_1+\int \delta \phi_2+\int \delta \phi_3) \nonumber \\
+e^{-a\int \delta\phi_2}(-2\int \delta \phi_2+\int \delta \phi_1+\int \delta \phi_3)
+e^{-a\int \delta\phi_3}(-2\int \delta \phi_3+\int \delta \phi_1+\int \delta \phi_2)\bigg] 
\end{eqnarray}

\section{Solving the Wheeler-DeWitt of small universe}\label{sec4}
By the same method we can quantize following universe 
\begin{eqnarray}
\begin{pmatrix}
-N^2 & 0 & 0 & 0 \\
0 & g_1(t,x,y,z) & 0 & 0 \\
0 & 0 & g_2(t,x,y,z) & 0 \\
0 & 0 & 0 & g_3(t,x,y,z) 
\end{pmatrix} \label{mat2}
\end{eqnarray}
Here $x,y,z$ are Gaussian normal coordinates.
The step 1 is to use a static restriction to the Eq.(\ref{eq2}), then we obtain
\begin{eqnarray}
\sum_i\frac{\delta^2}{\delta \phi_i^2}+2\phi_{i,jj}e^{\phi_i}=0. \label{eq4}
\end{eqnarray}
Then we use parameter separation, i.e. the solution of the above equation is assumed to be written as
\begin{eqnarray}
f[\phi_i]=f_1[\phi_1]f_2[\phi_2]f_3[\phi_3]
\end{eqnarray}
Then Eq. (\ref{eq4}) becomes
\begin{eqnarray}
\frac{\delta^2}{\delta\phi_i^2}+2\phi_{i,jj}e^{\phi_i}=0
\end{eqnarray}
Or,
\begin{eqnarray}
\frac{\delta^2}{\delta a_i^2}+8\partial_j\partial^j\ln a_i=0 \label{eq5}.
\end{eqnarray}
Here $a_i=g_i^{1/2}$.
Then we use a following equation
\begin{eqnarray}
-i(\ln a_{i,jj})^{1/2}\frac{\delta}{\delta a_i}+i\frac{\delta}{\delta a_i}(\ln a_{i,jj})^{1/2}=i\delta \frac{1}{2}(\ln a_{i,jj})^{-1/2}a_{i,jj}^{-1}
\end{eqnarray}
Now we briefly write $\partial_j\partial^j\ln a_i=\ln a_{i,jj}$.
Off course it is diferent, we simplisitily use the latter.
Then Eq.(\ref{eq5}) becomes
\begin{eqnarray}
\frac{\delta^2}{\delta a_i^2}+2\sqrt{2}i(\ln a_{i,jj})^{1/2}\frac{\delta}{\delta a_i}
-2\sqrt{2}i\frac{\delta}{\delta a_i}(\ln a_{i,jj})^{1/2}
+8\ln a_{i,jj}=i\sqrt{2}\delta (\ln a_{i,jj})^{-1/2}a_{i,jj}^{-1}
\end{eqnarray}
Or 
\begin{eqnarray}
\bigg( \frac{\delta}{\delta a_i}+2\sqrt{2}i(\ln a_{i,jj})^{1/2}\bigg)
\bigg( \frac{\delta}{\delta a_i}-2\sqrt{2}i(\ln a_{i,jj})^{1/2}\bigg)
=\sqrt{2}i\delta (\ln a_{i,jj})^{-1/2}a_{i,jj}^{-1}.
\end{eqnarray}
Then we take following assumptions 
\begin{eqnarray}
\frac{\delta}{\delta a_i}+2\sqrt{2}i(\ln a_{i,jj})^{1/2}=g_1[a_i] \\
\frac{\delta}{\delta a_i}-2\sqrt{2}i(\ln a_{i,jj})^{1/2}=g_2[a_i] .
\end{eqnarray}
Here,
\begin{eqnarray}
g_1[a_i]g_2[a_i]=i\delta \sqrt{2}(\ln a_{i,jj})^{-1/2}a_{i,jj}^{-1} \label{eq6}
\end{eqnarray}
Then second partial functional derivative become ordinal functional derivative and the equation become following 
\begin{eqnarray}
\frac{\delta f^{1/2}[a_i]}{\delta a_i}+2\sqrt{2}i(\ln a_{i,jj})^{1/2}f^{1/2}[a_i]=g_1[a_i]f^{1/2}[a_i] \\
\frac{\delta f^{1/2}[a_i]}{\delta a_i}-2\sqrt{2}i(\ln a_{i,jj})^{1/2}f^{1/2}[a_i]=g_2[a_i]f^{1/2}[a_i] .
\end{eqnarray}
From this equation we obtain
\begin{eqnarray}
\frac{1}{2}\ln f[a_i]=\int g_1[a_i]+ 2\sqrt{2}i(\ln a_{i,jj})^{1/2}\delta a_i \\
\frac{1}{2}\ln f[a_i]=\int g_2[a_i]- 2\sqrt{2}i(\ln a_{i,jj})^{1/2}\delta a_i
\end{eqnarray}
We can obtain the solution
\begin{eqnarray}
f[a_i]=\exp \bigg( 2\int g_1[a_i]+ 2\sqrt{2}i(\ln a_{i,jj})^{1/2}\delta a_i\bigg) \\
f[a_i]=\exp \bigg( 2\int g_2[a_i]- 2\sqrt{2}i(\ln a_{i,jj})^{1/2}\delta a_i\bigg) \label{eq7}
\end{eqnarray}
Because $f[a_i]$ is same 
\begin{eqnarray}
g_1[a_i]=g_2[a_i]-4\sqrt{2}(\ln a_{i,jj})^{1/2}.
\end{eqnarray}
Inserting this equation to Eq.(\ref{eq6}), we obtain 
\begin{eqnarray}
g_2[a_i]^2-4\sqrt{2}(\ln a_{i,jj})^{1/2}g_2[a_i]=i\delta \sqrt{2}(\ln a_{i,jj})^{-1/2}a_{i,jj}^{-1}.
\end{eqnarray}
Exchangeng this equation, we obtain
\begin{eqnarray}
g_2[a_i]^2-4\sqrt{2}(\ln a_{i,jj})^{1/2}g_2[a_i]-i\delta \sqrt{2}(\ln a_{i,jj})^{-1/2}a_{i,jj}^{-1}=0.
\end{eqnarray}
By solving this second order function equation, we obtain 
\begin{eqnarray}
g_2[a_i]=2\sqrt{2}(\ln a_{i,jj})^{1/2}\pm \sqrt{8\ln a_{i,jj}+
i\delta \sqrt{2}(\ln a_{i,jj})^{-1/2}a_{i,jj}^{-1}}.
\end{eqnarray}
Inserting this equation to Eq.(\ref{eq7}), we obtain
\begin{eqnarray}
f[a_i]=\exp \bigg( \int \pm 2\sqrt{8\ln a_{i,jj}
+i\delta \sqrt{2}(\ln a_{i,jj})^{-1/2}a_{i,jj}^{-1}}\delta a_i\bigg) .
\end{eqnarray}
So the quantization of the spacetime as (\ref{mat2}) is
\begin{eqnarray}
f[a_1,a_2,a_3]=\prod_i\exp \bigg( \int \pm 2\sqrt{8\ln a_{i,jj}
+i\delta \sqrt{2}(\ln a_{i,jj})^{1/2}a_{i,jj}^{-1}}\delta a_i\bigg) .
\end{eqnarray}
This is the special solution of quantization of the spacetime (\ref{mat2}).

Then we carry out step 2, we remove the static restriction and we assume the state is 
the form of the
\begin{eqnarray}
|\Psi\rangle =f[a_i]h[a_i]
\end{eqnarray}
Then we obtain 
\begin{eqnarray}
\nabla (\nabla +h'[a_i])h[a_i]=0.
\end{eqnarray}
Here,
\begin{eqnarray}
h'[a_i]=\sum_i\pm \sqrt{8\ln a_{i,jj}+i\delta \sqrt{2}(\ln a_{i,jj})^{1/2}a_{i,jj}^{-1}}
\end{eqnarray}
The functional partial derivative 
\begin{eqnarray}
(\nabla +h'[a_i])h[a_i]=0
\end{eqnarray}
is solved analytically.
The solution is 
\begin{eqnarray}
h[a_i]=\exp \bigg( \int^{a_1} h'[a_1',-a_1+a_2+a_1',-a_1+a_3+a_1']\delta a_1'\bigg) (-2\int \delta a_1+\int \delta a_2+\int \delta a_3) \nonumber \\
+\exp \bigg( \int^{a_2} h'[a_1-a_2+a_2',a_2',-a_2+a_3+a_2']\delta a_2'\bigg) (-2\int \delta a_2+\int \delta a_1+\int \delta a_3) \nonumber \\
+\exp \bigg( \int^{a_3} h'[-a_3+a_1+a_3',-a_3+a_2+a_3',a_3']\delta a_3'\bigg) (-2\int \delta a_3+\int \delta a_1+\int \delta a_2)
\end{eqnarray}
So the quantum state of spacetime (\ref{mat2}) is
\begin{eqnarray}
|\Psi\rangle =\prod_i\exp \bigg( \int \pm 2\sqrt{8\ln a_{i,jj}
+i\delta \sqrt{2}(\ln a_{i,jj})^{-1/2}a_{i,jj}^{-1}}\delta a_i\bigg) \nonumber \\
\times \bigg[ \exp \bigg( \int^{a_1} h'[a_1',-a_1+a_2+a_1',-a_1+a_3+a_1']\delta a_1'\bigg) (-2\int \delta a_1+\int \delta a_2+\int \delta a_3) \nonumber \\
+\exp \bigg( \int^{a_2} h'[a_1-a_2+a_2',a_2',-a_2+a_3+a_2']\delta a_2'\bigg) (-2\int \delta a_2+\int \delta a_1+\int \delta a_3) \nonumber \\
+\exp \bigg( \int^{a_3} h'[-a_3+a_1+a_3',-a_3+a_2+a_3',a_3']\delta a_3'\bigg) (-2\int \delta a_3+\int \delta a_1+\int \delta a_2) \bigg]
\end{eqnarray}
This is the main result of our paper.
\section{Conclusion and Discussions}\label{sec5}
We quantized diagonal metric space with Gaussian normal coordinate.
By solving the Wheeler-DeWitt equation,
we know the form of the solution.
We quantize two universe.
The common feature is similarity of step 2.
The step 2 is all ways solved.
So the important point is to find the special solution. 

There are many further work.
Once we should calculate the inner product and norm.
If the normalization is end the state is used to calculate the averaged value.
Or we should discuss problem of the norm.
This open issue is very important one.
And we should search the initial singularity or the black hole singularity.
Our work does not end. 
 
\end{document}